\documentclass[reprint,aps,prb,superscriptaddress]{revtex4-2}

\usepackage[T1]{fontenc} 
\usepackage{bm,physics,amsmath,amssymb,
	graphicx,xcolor,embedfile,comment,mathrsfs,mathtools,xfrac}
\usepackage{color,soul}
\usepackage[utf8]{inputenc}
\usepackage[normalem]{ulem}
\usepackage{nicefrac, xfrac}
\usepackage[T1]{fontenc}
\usepackage{tikz}
\usepackage{chemformula} 

\usepackage[mathscr]{euscript}

\DeclareSymbolFont{rsfs}{U}{rsfs}{m}{n}
\DeclareSymbolFontAlphabet{\mathscrsfs}{rsfs}

\graphicspath{{figures/}}

\usepackage[english]{babel}
\makeindex

\begin{document}

\title{Bond polarizability as a probe of local crystal fields in hybrid lead-halide perovskites}

\author{Yujing Wei}
\altaffiliation{Current address: Department of Chemistry, Columbia University, New York, New York 10027, United States}
\affiliation{Institute of Science and Technology Austria (ISTA), Am Campus 1, 3400 Klosterneuburg, Austria}

\author{Artem G. Volosniev}
\affiliation{Institute of Science and Technology Austria (ISTA), Am Campus 1, 3400 Klosterneuburg, Austria}

\author{Dusan Lorenc}
\affiliation{Institute of Science and Technology Austria (ISTA), Am Campus 1, 3400 Klosterneuburg, Austria}

\author{Ayan A. Zhumekenov}
\altaffiliation{Current address: School of Materials Science and Engineering, Nanyang Technological University, 50 Nanyang Avenue, 639798 Singapore}
\affiliation{KAUST Catalysis Center (KCC), Division of Physical Sciences and Engineering, King Abdullah University of Science and Technology (KAUST), Thuwal 23955-6900, Kingdom of Saudi Arabia}
    
 \author{Osman M. Bakr}
\affiliation{KAUST Catalysis Center (KCC), Division of Physical Sciences and Engineering, King Abdullah University of Science and Technology (KAUST), Thuwal 23955-6900, Kingdom of Saudi Arabia}    
    
\author{Mikhail Lemeshko}
\affiliation{Institute of Science and Technology Austria (ISTA),
    Am Campus 1, 3400 Klosterneuburg, Austria} 

\author{Zhanybek Alpichshev}
\email{alpishev@ist.ac.at}
\affiliation{Institute of Science and Technology Austria (ISTA), Am Campus 1, 3400 Klosterneuburg, Austria}

\begin{abstract}
\noindent A rotating organic cation and a dynamically disordered soft inorganic cage are the hallmark features of hybrid organic-inorganic lead-halide perovskites. Understanding the interplay between these two subsystems is a challenging problem but it is this coupling that is widely conjectured to be responsible for the unique behavior of photo-carriers in these materials. In this work, we use the fact that the polarizability of the organic cation strongly depends on the ambient electrostatic environment to put the molecule forward as a sensitive probe of local crystal fields inside the lattice cell. We measure the average polarizability of the C/N--H bond stretching mode by means of infrared spectroscopy, which allows us to deduce the character of the motion of the cation molecule, find the magnitude of the local crystal field and place an estimate on the strength of the hydrogen bond between the hydrogen and halide atoms. Our results pave the way for understanding electric fields in lead-halide perovskites using infrared bond spectroscopy.
\end{abstract}



\maketitle

\newpage

\noindent The efficiency of hybrid organic-inorganic lead-halide perovskite (HOIP)-based solar cells has recently nearly reached the levels of the state-of-the-art conventional Si-based devices \cite{Min2021,Almora2022}. On the microscopic level this impressive performance is contingent upon the presence of charged photocarriers that can travel unimpeded sufficiently far to reach the contacts of the photo-voltaic element. The fact that this is indeed the case for lead-halide perovskites \cite{Dong2015,Turedi2022,Shi2015, Xing2013, Brenner2016, Stranks2013}, despite much higher defect concentration in HOIP samples as compared to standard photo-voltaic materials such as Si~\cite{Lekesi2022}, is arguably the biggest puzzle in the field of perovskite research. While an exhaustive explanation is still lacking, the unexpectedly high efficiency of HOIPs, be it in photo-carrier separation \cite{Miyata2017} or `neutralization' of defect centers \cite{Yin2014}, suggests the pre-eminent role of charge screening. One possible source of such screening pointed out early on is the quasi-freely rotating polar organic cation occupying the A-site of HOIPs such as CH$_3$NH$_3^+$ (methylammonium, MA) \cite{Zhu2015, Zhu2016, Frost2014,Koutentakis2023}, while others argue for the inorganic cage as the main responsible party \cite{Kang2017}. In order to bring more clarity to this problem, a detailed understanding of the interaction between cation molecules and the surrounding cage is necessary. Unsurprisingly, this is not an easy task as while the former is semi-independent, the latter is soft, dynamically disordered  and highly anharmonic \cite{Leguy2016, Ferreira2018, Ferreira2020, LaniganAtkins2021, Yaffe2017, Songvilay2019} with significant ionic conductivity \cite{Peng2018} to the extent of being dubbed ``plastic crystals'' in some works~\cite{Whalley2017,Mozur2021}). 

\begin{figure*}
\begin{center}
    \includegraphics[width=\textwidth]{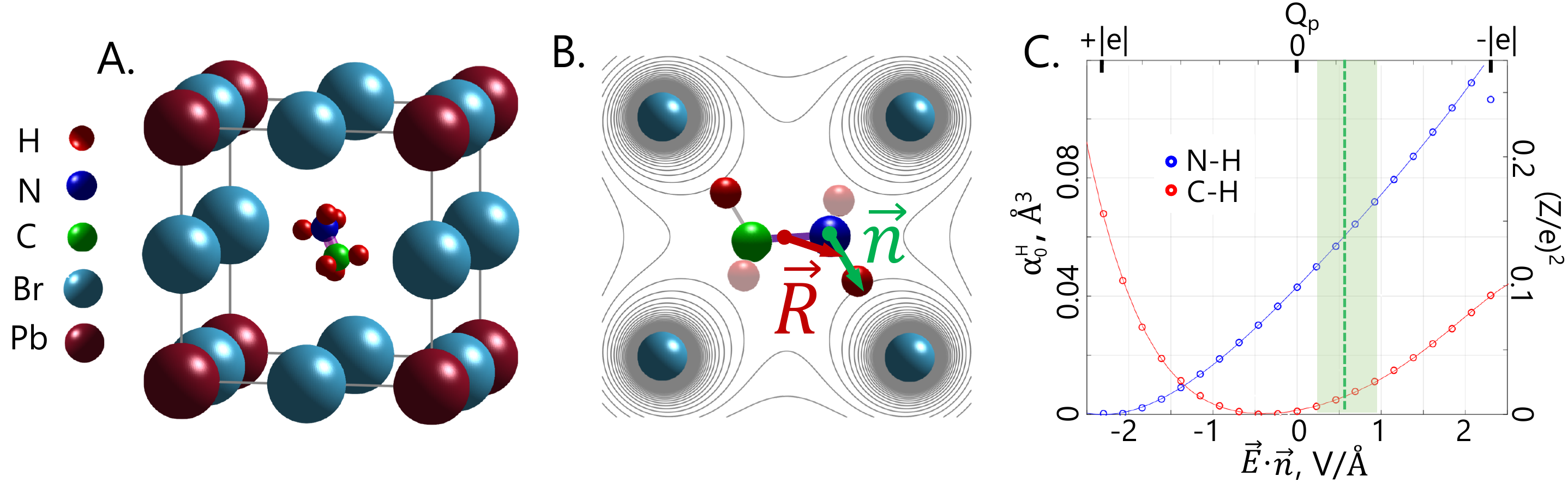}
\end{center}
\caption{(A) Lattice structure of MAPbBr$_3$ (to scale). (B) Cartoon of the MA cation in the electrostatic field of the inorganic cage (not to scale); $\vec{n}$ is the unit vector along the bond direction; $\vec{R}$ is the distance to the center of the bond from the center of the lattice unit. (C) Static longitudinal polarizabilities $\alpha^{H}_0$ of C(N)--H bonds as a function of the electric field at the locus of the bond. For comparison, the field is also parametrized by the value of a point charge $Q_p$ placed on the line connecting one of the H atoms to the neighboring N(C) atom, at a distance $d=2.5$ \r{A} from H. This $d$ value corresponds to the most stable H--Br separation in the cubic phase of MAPbBr$_3$ for the N--H hydrogen~\cite{Yin2017,Varadwaj2018} (for consistency, the same distance is chosen for C--H in our calculations). $\alpha^{H}_0$ is also parametrized with a square of the effective (Born) charge of an H atom, $Z$, as $\alpha^{H}_0 = Z^2/m_H \Omega_H^2$ (see~\cite{supplementary}), where $m_H$ is the mass of a hydrogen atom, and $\Omega_H$ is the resonant frequency of the C(N)--H stretching mode. The green dashed line (green rectangle) marks the estimated value (error margin) of the crystal field experienced by the C(N)--H bond (see the text for details).}
  \label{fig:DFT}
\end{figure*}

The very fact that the A-site cation retains its chemical autonomy inside the cage~\cite{Filippetti2014,Perez-Osorio2015,Mattoni2017}, implies that the interaction between the two is mostly of electrostatic nature~\cite{Govinda2017}. To characterize the magnitude of this coupling one needs therefore to be able to probe the local electric fields within the crystal, primarily the ones experienced by hydrogen atoms. Unfortunately, conventional tools such as nuclear quadrupole resonance spectroscopy~\cite{Senocrate2018,Piveteau2020a,Piveteau2020b} are not particularly suitable for this purpose as the lightest nuclei (proton, deuteron) have either no or small quadrupolar moment.  
In this light we draw attention to the fact that in HOIPs, the A-site cation can itself act as a sensitive local probe of its immediate environment. Indeed, on the one hand the cation is chemically decoupled from the inorganic cage and can be treated as a separate molecule, and on the other hand, it is often simple enough to allow for an exhaustive theoretical treatment.  

In this work, we show that the ambient electrostatic environment can noticeably modify basic properties of the molecule such as its dynamic susceptibility, using methylammonium in MAPbBr$_3$ as an example. We compare these theoretical results to the measured infrared polarizability of methylammonium, and draw conclusions about the character of coupled cation-cage dynamics, put a numeric value on the effective local electric field experienced by the hydrogen atoms in methylammonium cation and estimate the strength of the hydrogen bond between H and Br atoms. These findings pave the way for employing infrared probes~\cite{Barth2007,Ma2015} for studying organic-inorganic crystal structures such as HOIPs.

Polarizability $\alpha_{ij}$ is a basic physical property of a molecule that relates the induced dipole moment, ${\vec p}$, of a molecule to the applied electric field, ${\vec E}$, as $p_i(\omega) = \alpha_{ij}(\omega) E_j(\omega)$. Microscopically, ${\vec p}$ can be induced either by the electron cloud polarization or by deformation of the molecule in question. When the field ${\vec E}$ becomes comparable to the inter-atomic one, the linear relation between ${\vec p}$ and ${\vec E}$ above is no longer valid and should be amended by assuming that $\alpha_{ij}$ depends on ${\vec E}$~\cite{Boyd1999}. Such fields are natural for intra-cell lattice environments, meaning that even if the molecule retains its chemical autonomy within a compound, one should expect its vibrational polarizability to be noticeably affected by the local fields inside a lattice unit. This turns the polarizability of the molecule into a marker of these fields, which can be investigated in linear optical experiments where probing fields are weak by definition.

Any molecular deformation can be decomposed into normal vibration modes \cite{Landau1976Mechanics}, where each normal mode $k$ has a specific resonance frequency $\Omega_k$ and can be ascribed a frequency-dependent polarizability tensor $\alpha^k_{ij}(\omega)$. The sum of these terms determines the refractive index $n(\omega)$ of the medium; the contribution of each normal mode to $n$ is the most prominent near its resonant frequency $\Omega_k$ \citep{Born2019}. In this work, we choose to focus on the vibration modes of methylammonium that correspond to the longitudinal displacement of the hydrogen atoms along the direction of the bond, $\vec{n}$, to the nearby N or C atoms (see Figs.~\ref{fig:DFT}A and~\ref{fig:DFT}B). The main reason for this choice is that these modes are by far the strongest in terms of IR-intensity~\cite{Leguy2016,Perez-Osorio2015,Perez-Osorio2018}.  Elementary counting of the degrees of freedom reveals (in total) 6 modes that involve longitudinal stretching of the C(N)--H bond, which will be labelled as $H_k$ or simply $H$. 

\begin{figure*}[t]
\begin{center}
    \includegraphics[width=\textwidth]{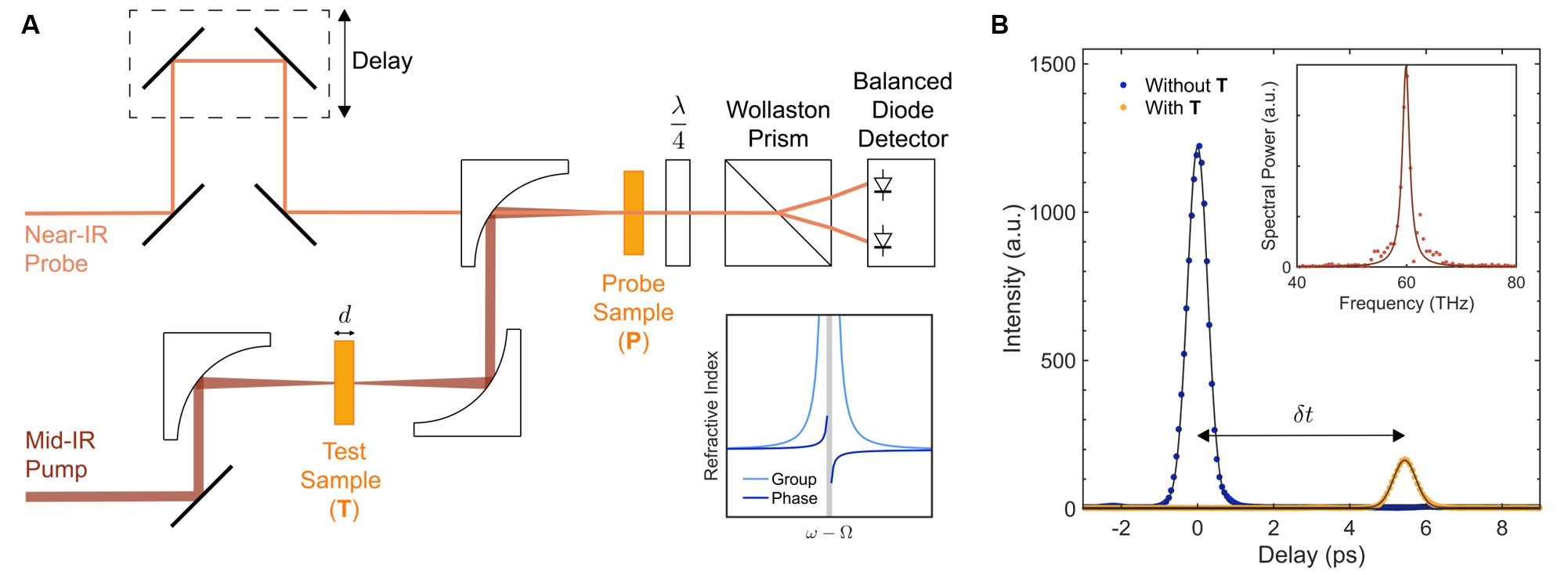}
\end{center}
\caption{(A) Optical Kerr effect-based pump-probe method to obtain the group refractive index by measuring the time it takes for a pulse to go through the test sample. The delay between the mid-IR pump and the near-IR probe is measured using balance detection of the mid-IR induced birefringence in the probe sample. The inset shows the region of anomalous dispersion for the phase and group refractive indexes, showing that the group refractive index is more divergent. (B) Example of measured signals of the pump-probe delay with and without test sample. The inset shows Fourier-transform infrared spectroscopy (FTIR) of the pump pulse.}
  \label{fig:Exp_Method}
\end{figure*}

The simplicity of methylammonium allows for a direct computation of the polarizability of the C(N)--H stretching mode, $\alpha^{H}_0$, by means of density functional theory (DFT); for more details see~\cite{supplementary}. In Fig.~\ref{fig:DFT}C we show how the (longitudinal) polarizabilities of the C--H and N--H bonds depend on the strength of the longitudinal electric field. To give these values some intuitive sense, the applied electric field at the locus of the C(N)--H bond is also parametrized by the charge $Q_p$ of a point source located on the C(N)--H line at a distance $d=2.5$~\r{A} from the H atom; the value for $d$ is motivated by the H--Br distance at room temperature~\cite{Yin2017,Varadwaj2018,Abia2022}. Although, the present work focuses on the stoichiometric case, we note that our calculations indicate that the effect of localized lattice defects such as vacancies or color centers must be very significant. Indeed, the associated fields are on the order of $E_{\mathrm{def}} \sim |e|/d^2 \sim 1$ V/\r{A} in the vicinity of the molecule. They can renormalize bond polarizability by about 100\%, meaning that methylammonium cation can be a very sensitive probe of lattice imperfections.

The microscopic longitudinal bond polarizability $\alpha^H_{ij}$ depends on the fluctuating local crystal field $\vec{e}({\vec R})$ (see Fig.~\ref{fig:DFT}C). To relate it to the polarizability $\langle \alpha^H \rangle$ measured in experiment, $\alpha^H_{ij}$ must be averaged over all spatial orientations of the bond and different configurations of $\{ \vec{e}({\vec R}) \}$. For MAPbBr$_3$, which is cubic (on average) at room temperature, one can write (see \cite{supplementary}):
\begin{equation}
\langle \alpha^H \rangle = \frac{1}{3} \sum_{H_k=1}^6 \int  d\vec{X} \, d\vec{Q} \,\rho(\vec{X},\vec{Q}) \, \alpha^{H_k}(\vec{X},\vec{Q}),
\label{eq:average_pol_density_matrix}
\end{equation}
where the index $H_k$ runs over the six longitudinal C(N)--H stretching modes~\footnote{Note that we assume that the polarizability of the molecule is the sum of polarizabilities of its bonds~\cite{Denbigh1940,LeFevre1965}}; $\vec{X}$ ($\vec{Q}$) are the positions of atoms in the molecule (the cage); $\rho(\vec{X},\vec{Q})$ is the density matrix that determines probabilities of a specific lattice-molecule configuration given by $\vec{X}$ and $\vec{Q}$; $\alpha^{H_k}(\vec{X},\vec{Q})$ is the longitudinal polarizability of the $H_k-$th bond in a configuration $\{ \vec{X}, \,\vec{Q}\}$ (here we rely on the Born-Oppenheimer approximation to assume that $\alpha^H(\vec{X},\vec{Q})$ only depends on instantaneous positions of atoms in the cage). To derive Eq.~(\ref{eq:average_pol_density_matrix}), we noted that according to our DFT calculations, $\alpha^H_{ij}(\vec{e}({\vec R})) = \alpha^{H_k}({\vec R}) \, n_i n_j$, where $\vec{n}$ is a unit vector along the bond direction, see~\cite{supplementary} for further details.

The quantities $\rho$ and $\alpha^H$ in Eq.~(\ref{eq:average_pol_density_matrix}) can be determined theoretically from the literature and DFT calculations for different scenarios, such as the presence of defects. Equation~(\ref{eq:average_pol_density_matrix}) relates these theoretical calculations to the measured average polarizability $\langle \alpha^H \rangle$, allowing one to investigate microscopic properties of HOIPs using molecules as experimental probes.

\begin{figure*}
\begin{center}
\includegraphics[width=\textwidth]{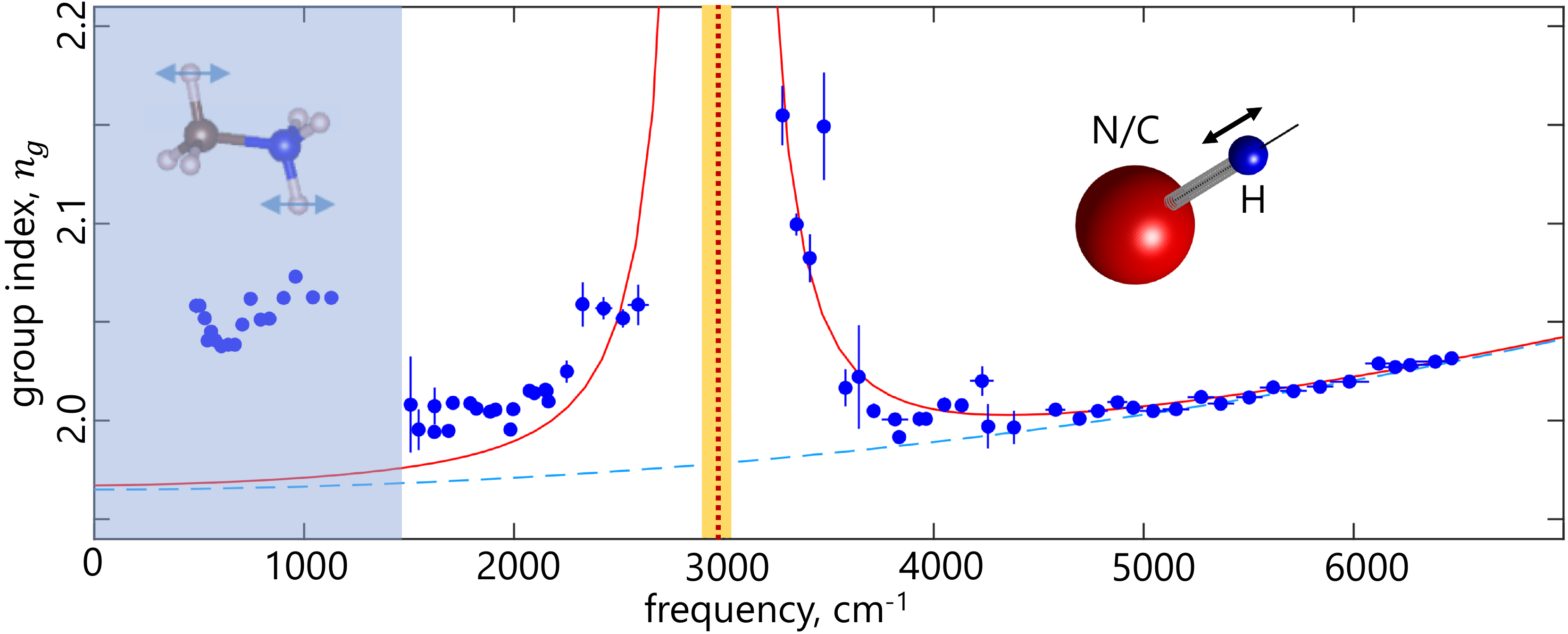}
\end{center}
\caption{Group refractive index $n_g$ measured in the setup depicted in Fig.~\ref{fig:Exp_Method}. The horizontal error bars are the full width half maximum (FWHM) of the FTIR calibration, and the vertical error bars are the FWHM of the dispersion from Fig.~\ref{fig:Exp_Method}B. The low-frequency region of vibrational resonances is indicated by the blue shading. Sketch of the corresponding (bending) molecular modes is presented in the shaded region. The right inset is an illustration of a longitudinal C(N)--H bond stretching, representing the entire cluster of 6 stretching modes in the energy region around 3000 cm$^{-1}$. The solid red line is the fit according to Eq.~(\ref{eq:group_mol}); the vertical red dotted line marks the resonant frequency, $\Omega_H$, from the fit; the light blue dashed curve is the electronic group index obtained from the Sellmeier fit to the high energy data of Ref.~\cite{ArtemArxiv2020}; the yellow shaded region illustrates the frequency spread within the longitudinal-stretching mode cluster~\citep{Leguy2016, Perez-Osorio2015, Schuck2018, Perez-Osorio2018}.}
\label{fig:Data_Fig}
\end{figure*}

To illustrate the formal discussion above, we proceed with measuring the average polarizability $\langle \alpha^H \rangle $ for a bulk single crystal sample of MAPbBr$_3$ in the cubic phase. For this purpose, we measure the group refractive index $n_g(\omega)$ in the mid-infrared frequency range relevant for vibrational degrees of freedom of methylammoium~\citep{Leguy2016}. The advantage of the group index for probing molecular polarizability -- which in general provides much weaker contribution to the refractive index as compared to the electronic one -- lies in the fact that $n_g(\omega)$ features stronger divergence near resonances and is therefore more sensitive as compared to the phase index $n_{ph}(\omega)$. Indeed, $n_g(\omega)/n_{ph}(\omega)\approx \Omega/\left(  \omega -\Omega \right)$ near the resonance frequency $\Omega$ (see the inset in Fig.~\ref{fig:Exp_Method}A). 

To measure the group refractive index in a direct manner, we develop a time-resolved setup depicted in Fig.~\ref{fig:Exp_Method}A. The setup utilizes two samples of single-crystal MAPbBr$_3$: the actual test sample (\textbf{T} in Fig.~\ref{fig:Exp_Method}) in which $n_g$ is measured, and a probe sample for a broadband ultrafast detection of mid-infrared pulses by means of the optical Kerr effect (\textbf{P} in Fig.~\ref{fig:Exp_Method}). By measuring the time of arrival of the mid-IR pulse at the probe sample with and without the test sample, one obtains the delay $\delta t$ introduced by the test sample. The time delay $\delta t$ for the mid-IR pulse due to the travelling through the perovskite sample of thickness $d=1.6$ mm, is $\delta t = \sfrac{d}{c} \left( n_g - n_g^{\text{air}} \right)$, where $n_g$ and $n_{g}^{\text{air}} \approx 1$ are the group refractive indices of MAPbBr$_3$ and air, respectively; $c$ is the speed of light. 

The ultrafast optical Kerr effect in the probe sample is detected by the standard time resolved pump-probe method in a balanced detection scheme (see Fig.~\ref{fig:Exp_Method}A and~\cite{supplementary}). An example of a time-resolved transient Kerr response for $\lambda=5$ $\mu$m mid-IR pump with and without the test sample is depicted in Fig.~\ref{fig:Exp_Method}B. The difference in arrival time of the mid-IR pulse at the probe sample $\delta t$ is measured as the distance between the two peak positions; the inset shows the FTIR of the pump pulse (see~\cite{supplementary}). Performing similar measurements for other mid-IR wavelength values, one can extract the perovskite group index $n_g(\omega) = n_g^\text{air}+ c\,\delta t /{d} $ in a broad range of wavelengths.

The measured group refractive index dispersion $n_g(\omega)$ of a bulk single crystal MAPbBr$_3$ as a function of the photon energy, $\hbar \omega$, is displayed in Fig.~\ref{fig:Data_Fig}. Here, we identify two regions of absorption, one in the low energy region at $\sim0.17$ eV (1400 cm$^{-1}$) and another at a higher energy $\sim 0.37$ eV (3000 cm$^{-1}$), which can be associated with the bending and stretching modes of C(N)--H bonds, respectively~\cite{Leguy2016,Perez-Osorio2015,Perez-Osorio2018}. The low value of extinction, $\kappa$, (see~\cite{supplementary}) in the region outside the immediate vicinity of the absorption band corresponding to the C(N)--H stretching modes justifies our assumption that these modes can be approximated as isolated from the nearby C(N)--H bending modes. 

To extract the molecular contribution to the refractive index, we need to subtract the electronic contribution from the measured infrared $n_g(\omega)$~\footnote{Note that the electronic contribution to the refractive index is typically the dominant one, since electrons are much lighter than ions. Only close to a molecular resonance, the molecular contribution becomes important.}. To this end, we use our previous measurements of the phase index in the visible and near-IR region \cite{ArtemArxiv2020}. In this high frequency range near the bulk band transition frequency, $\hbar \omega \sim \Delta_\text{gap}$ and the refractive index is determined by the electronic polarizability of the lead-halide cage. By first fitting the high-frequency phase index from Ref.~\cite{ArtemArxiv2020} with a Sellmeier expression, then calculating group index from it and extrapolating the result to the mid-IR frequencies, one can obtain the electronic part $n_{g,el}(\omega)$ of the total group refractive index $n_g(\omega)$ coming from the crystal cage. $ n_{g,el}(\omega)$ in the mid-IR range calculated this way is plotted as a light blue dashed line in Fig.~\ref{fig:Data_Fig} providing an excellent fit to the data in the relevant frequency range away from the nearby absorption band. The molecular part of the group index $n_{g,mol}$ is found by subtracting the cage electronic contribution from the total refractive index: $n_{g,mol}(\omega)=n_g(\omega)- n_{g,el}(\omega)$ (see~\cite{supplementary}). 

\par Once $ n_{g,mol}(\omega)$ is known, it is possible to calculate the polarizability of the given mode by fitting it to the classical  resonant profile near the resonant frequency. We focus on the C(N)--H stretching modes centered around $\hbar\Omega_H\approx0.37$ eV (3000 cm$^{-1}$). Note that the resonance in  Fig.~\ref{fig:Data_Fig} corresponds not to a single mode but instead to a group of 6 different modes involving stretchings of different C(N)--H bonds. However, due to strong mass mismatch between H and C(N), the natural frequencies of these modes are very similar~\cite{Perez-Osorio2015, Leguy2016, Perez-Osorio2018}. If we assume that C and N are infinitely heavy compared to hydrogen, then each vibration of H along the C(N)--H bond can be considered as independent and decoupled from the rest. Within this approximation (that introduces an error of the order $\sim m_H/m_C\approx 10\%$), we shall use an oscillator model with a single resonant frequency (see the inset in Fig.~\ref{fig:Data_Fig}).

To connect $ n_{g,mol} (\omega)$ to $\langle\alpha^H_{0}\rangle$, we account for the screening effect due to the polarizable cage and write in the vicinity of $\Omega_H$ (see~\cite{supplementary} for derivation),
\begin{equation}
    n_{g,mol} (\omega) \approx \langle \alpha^H_{0}\rangle \frac{4\pi N}{3}\frac{(\bar{n}_{el}^2+2)^2}{12\bar{n}_{el}} \cdot \frac{\Omega_H^2}{(\Omega_H-\omega)^2},
\label{eq:group_mol}
\end{equation}
\noindent where $\bar{n}_{el}\equiv  n_{ph,el}(\Omega_H)$ is the electronic contribution of the phase refractive index at $\Omega_H$, and $\langle \alpha^H_{0}\rangle$ is the average static (longitudinal) polarizability of the C(N)--H bond.  The solid line in Fig.~\ref{fig:Data_Fig} is the result of the fit of $ n_{g,mol} (\omega)$  to the expression in Eq.~(\ref{eq:group_mol}). From the fit, we obtain $\langle \alpha^H_{0} \rangle \approx (7.0\pm1.1)\times 10^{-2}$ \r{A}$^3$, error representing the 95\% confidence range. Since the fitting error significantly exceeds the spread in resonant frequency values for C(N)--H stretching modes ($\sim 10 \%$, see above) one can neglect the latter for the sake of discussion.

With an experimental value for $\langle \alpha^H \rangle$ at hand,  we compare it with Eq.~(\ref{eq:average_pol_density_matrix}) to see what can be said about the interaction between MA and the surrounding cage. In the simplest case, one may note that in some molecular dynamics studies the C--N axis of the cation appears to explore the available space quasi-uniformly~\cite{Mattoni2015,Gallop2018} and naively assume that the molecule rotations are uncorrelated with the inorganic cage. In terms of Eq.~(\ref{eq:average_pol_density_matrix}), this limit corresponds to $\rho(\vec{X},\vec{Q})=\rho_{\mathrm{mol}}(\vec{X})\rho_{\mathrm{cage}}(\vec{Q})$, for which (see~\cite{supplementary}) the average polarizability remains un-renormalized even in the presence of strong crystal fields $\langle \alpha^H_0 \rangle_{\mathrm{uncorr}} = \alpha^{H_N}_0(\vec E=0)+\alpha^{H_C}_0(\vec E=0)$, where $\alpha^H_0(E=0)$ can be taken from Fig.~\ref{fig:DFT}C so that $\langle \alpha^H_{0} \rangle_{\mathrm{uncorr}} \approx 4.4\times 10^{-2}$ \r{A}$^3$. Since the discrepancy between this value and the experimental one above is beyond the error margin of our approach, we can conclusively rule out uncorrelated molecule-cage dynamics, in agreement with prior results~\cite{Leguy2015,Yin2017,Svane2017,Saleh2021}. If, conversely, we fix the density matrix in Eq.~(\ref{eq:average_pol_density_matrix}) in accordance with previous reports in literature~\cite{supplementary}, then we arrive at the value for local crystal field experienced by the N--H bond $|E_{\mathrm{CF}}| = (0.6\pm 0.3)$ V/\r{A} by comparing the polarizabilities in Fig.~\ref{fig:DFT}C with experimental values.

Note that in a stoichiometric lattice the Coulombic fields coming from the atoms, if treated as point sources, cancel each other very finely inside the lattice cell, see~\cite{supplementary}. Therefore, to account for the relatively large value of $E_{\mathrm{CF}}$, we need to go beyond the point-source approximation for the atoms in the inorganic cage. The lowest-order multipole moment would be the quadrupolar moment $D$ of the bromine atom, with the axis along Pb--Br--Pb. Knowing $|E_{\mathrm{CF}}|$ and the distance between H and Br one can estimate that $D \sim +1$ e\r{A}$^2$. Recalling now, that a hydrogen bond can be defined as an interaction between H and a multipole potential of a neighboring atom \cite{Arunan2011}, we estimate it from the energy of the H--Br electrostatic interaction $\mathcal{E}_H \sim 0.1$ eV (for more details, see~\cite{supplementary}). This value is in reasonable agreement with previous estimates~\cite{Lee2016, Svane2017} of the energy of the H--Br hydrogen bond, illustrating that the proposed here molecular probe can provide an important insight into the existence of hydrogen bonding in HOIPs~\cite{Glaser2015,Ibaceta2022}.

In conclusion, we propose the semi-autonomous A-site cations in HOIPs as a sensitive local probe for optical spectroscopy, which complements the existing techniques such as NMR and NQR. To illustrate the formulated theoretical framework, we have analysed the average polarizability of the N--H stretching mode extracted by measuring the group refractive index of a bulk single crystal sample of MAPbBr$_3$ in the mid-IR wavelength range. Based on the analysis, we have ruled out the possibility of an uncorrelated motion of methylammonium cation in a PbBr cage and estimated local electric fields at the locus of the C(N)--H bond. We also estimated the value of the quadrupole moment for the Br atom, and the energy of the H--Br hydrogen bond. Our work proposes a new approach to the study of the complex behavior of the dynamically disordered inorganic cage, which will provide novel insight into fundamental properties of lead-halide perovskites. In particular, being all-optical, our approach can be employed to study ultrafast transient behavior of lattice irregularities, for example polaronic structures formed around photo-excitations. Since it is widely expected that the formation of such polarons underlies some most important optoelectronic properties of lead halide perovskites~\cite{Miyata2017}, our results offer an approach to clarify some of the most pressing questions related to the solar energy harvesting applications of lead-halide perovskites.

\begin{acknowledgements}
We thank Bingqing Cheng and Hong-Zhou Ye for valuable discussions; Y.W.'s work at IST Austria was supported through ISTernship summer internship program funded by OeAD-GmbH; D.L. and Z.A. acknowledge support by IST Austria (ISTA); M.L. acknowledges support by the European Research Council (ERC) Starting Grant No.801770 (ANGULON). A.A.Z. and O.M.B. acknowledge support by KAUST.
\vspace{1em}

Y.W. and A.G.V. contributed equally.
\end{acknowledgements}

\newpage

\widetext
\section{Supplementary Information}

\subsection{Sample Preparation}

High quality bulk single crystal samples of MAPbBr$_3$ were grown by the inverse temperature crystallization method as described previously~\cite{Saidaminov2015,Saidaminov2015_2}.

{\it Chemicals.---} CH$_3$NH$_3$Br ($>$99.99\%) was purchased from GreatCell Solar Ltd. (formerly Dyesol) and used as received. PbBr$_2$  ($\geq$98\%),  DMF  (anhydrous,  99.8\%),  and  DMSO  (anhydrous, $\geq$99.9\%)  were  purchased  from  Sigma Aldrich  and  used  as  received. 

{\it Synthesis of MAPbBr$_3$ perovskite single crystals.---}A 1.5 M solution of CH$_3$NH$_3$Br/PbBr$_2$ in DMF was prepared, filtered through 0.45-$\mu$m-pore-size PTFE filter; and the vial containing $0.5-1$ ml of the solution was placed on a hot plate at $30^{\circ}$C. Then the solution was gradually heated to $\sim\!60^{\circ}$C and kept at this temperature until the formation of MAPbBr$_3$ crystals. The crystals can be grown into larger sizes by elevating the temperature further. Finally, the crystals were collected and cleaned using a Kimwipe paper.

\subsection{Calculation of bond polarizability}

\noindent \textbf{DFT Methods.}
The Born charges of H ions in the methylammonium were calculated by density functional theory (DFT) using the ORCA software package~\cite{neese2012orca}. The parameters used for the calculation are as follows: B3LYP functional~\cite{becke1993,lyp1988} and aug-cc-pVTZ~\cite{dunning1989a,kendall1992a} basis set for all steps of the calculations.

\noindent \textbf{Bond polarizability.} Since molecular polarizability is a dynamic property and every vibrational mode can be assigned its own contribution polarizability tensor, the focus of this work is on the most polarizable and therefore the most relevant modes. In the case of the methylammonium cation in the A-site of CH$_3$NH$_3$PbBr$_3$ hybrid lead-halide perovskite, by far the strongest modes in terms of IR-intensity are those that correspond to displacement of the hydrogen atoms along the direction of the bond to the nearby C or N atom \cite{Leguy2016,Perez-Osorio2015,Perez-Osorio2018}. Elementary counting of the degrees of freedom reveals that there are 6 modes that involve longitudinal stretching of the C(N)--H bond, but since both C and N atoms are much heavier than a proton, all of these modes are almost degenerate: $\Delta \Omega/\Omega_0 \sim m_H/m_C \approx 10\%$ is the error induced to our harmonic oscillator model with one resonant frequency $\Omega_H$. We calculate the polarizability of such modes using the Born effective charge. The relationship is as follows (see the discussion around Eq.~(\ref{eq:alpha_H_SM}) for more details):

\begin{equation}
\alpha_0^H=\frac{Z^2}{m_H\Omega_H^2}
\end{equation}

\noindent where $\alpha^{H}_0$ is the static longitudinal polarizability of C(N)--H bonds, $\Omega_H$ is the resonant frequency of the relevant cluster of modes that correspond to longitudinal bond stretching, and $Z$ is the Born effective charge, which is relatively simple to compute -- we refer to previous studies~\cite{Gough1989} that calculate polarizability in a similar way. 

To remind, Born effective charge $Z$ of a given ion is a tensor defined as
\begin{equation}
Z_{ij} = \frac{d p_i}{d r_j},
\end{equation}

\noindent where $p_i$ is the total dipole moment of the molecule and $r_j$ is the coordinate of the ion in question. In our case the polarizability of the C(N)--H bond stretching is related to the specific terms of the full Born charge tensor that correspond to the displacements of H ions along the corresponding bonds. In this work we calculate this ``longitudinal'' Born effective charge by following its definition above. Specifically, we first begin by finding the equilibrium atomic positions in the CH$_3$NH$_3^+$ ion by performing an initial geometry optimization and calculating the dipole moment of the resultant configuration. Then we displace the H ion in question by $\delta a = 0.01 a_0$ in the direction of C(N), where $a_0$ is the equilibrium C(N)--H bond length. In the new configuration we re-calculate the total dipole moment $\textbf{p}$ of the distorted molecule. An important detail here is that since net charge of methylammonium is not zero, we take care to keep the origin unmoved relative to the unperturbed atomic positions when re-calculating $\textbf{p}$. Provided the origin is kept fixed, the change $\delta \textbf{p} = \textbf{p} - \textbf{p}_0$ does not depend on it, and the Born effective charge corresponding to the ``longitudinal'' displacement of H along the C(N)--H bond is found as $Z=\delta \textbf{p} / \delta a$. 

\begin{figure}[h!]
\begin{center}
    \includegraphics[width=0.5\textwidth]{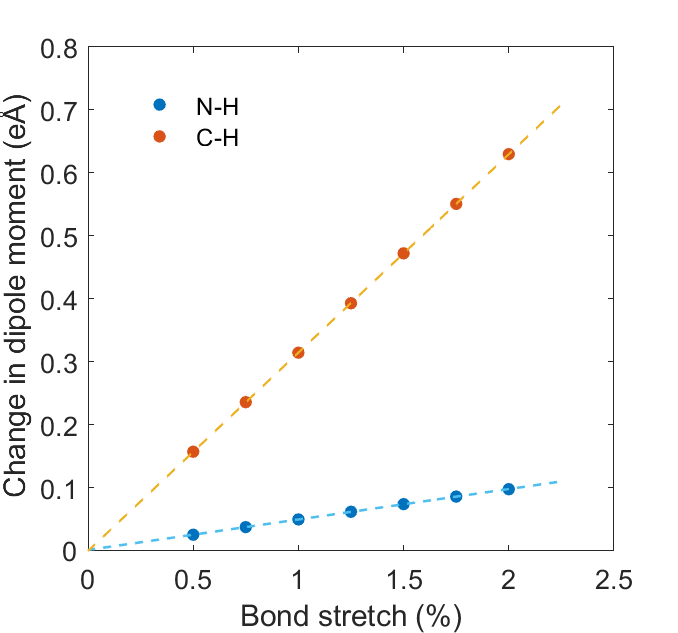}
\end{center}
\caption{Difference in dipole moment of methylammonium molecule upon a slight stretch in one C(N)-H bond from the equilibrium geometry. The bond stretch is presented as a percentage of the equilibrium C(N)-H bond length $a_0$ and is proportional to $\delta a$. The dashed lines are linear fits to the data, where the coefficient of determination $R^2=1.00$ for both N-H and C-H stretch.}
  \label{SI_linplot}
\end{figure}

In order to confirm that the chosen distortion $\delta a$ is appropriate for determining $Z$, we plot in Fig.~\ref{SI_linplot} the change of the dipole moment as a function of $\delta a$. We observe a linear dependence within the selected range, which confirms that $Z$ is a well-defined physical property.

\subsection{Derivation of Eq. (1) of the main text}
In an external field, the dipole moment of a bond in a molecule changes as
\begin{equation}
\Delta p_{\parallel}=\alpha_{\parallel\perp}E_{\perp} + \alpha_{\parallel\parallel}E_{\parallel}; \qquad
\Delta p_{\perp}=\alpha_{\perp\perp}E_{\perp} + \alpha_{\perp\parallel}E_{\parallel}.
\end{equation}
 Here, $\parallel$ and $\perp$ determine the directions with respect to the C(N)--H axis of the bond. Note that there are two directions perpendicular to the bond -- the notation $\perp$ is used for both, for convenience. We are interested in the polarizability $\alpha_{\parallel\parallel}$, which is denoted as $\langle \alpha^H_0\rangle$ in the main text. 
 
To relate $\langle \alpha^H_0\rangle$ to the density matrix $\rho$ (cf. Eq. (1) of the main text), we set $\alpha_{\parallel\perp}$ to zero, in agreement with our DFT calculations.
[To have a physical picture of this observation, one should consider the C(N)--H bond as a charged ellipsoid, see, e.g.,~\cite{LeFevre1965}.] Then, we
write the `longitudinal' part of the dipole moment of the molecule that is induced by a weak electric field (assuming that the center of mass is not changed):
\begin{equation}
\Delta {\vec{p}}_{\parallel}=\int d\vec{X} d\vec{Q}\rho(\vec{X},\vec{Q}) \sum_{H_k} \vec{n}_{k}(\vec{X},\vec{Q}) Z_{k}(\vec{X},\vec{Q}) \Delta X_{k},
\end{equation}
where the sum is over all C-H and N-H bonds; $\vec{n}_H$ is the unit vector that determines the direction of the bond, $Z$ is the associated charge. The change in the position of the $H_k$ atom along the bond, $\Delta X_{k}$, is calculated using the harmonic approximation 
\begin{equation}
\frac{\partial \left(m_H \Omega_H^2 X_k^2/2-X_k Z_k(\vec{E}\cdot \vec{n}_k)\right)}{\partial X_k}=0\;\to\;\Delta X_k=\frac{Z_k}{m_H\Omega_H^2}\vec{E}\cdot \vec{n}_k,
\end{equation}
where $m_H$ is the mass of a hydrogen atom. Note that we have assumed here that C and N atoms are infinitely heavy. This is done in line with the set of approximations made during the analysis of the data in the main text.

Let us now consider the quantity $\Delta \vec{p}_{\parallel}\cdot \vec{E}$. As the HOIP has (on average) a cubic lattice, $\Delta \vec{p}_{\parallel}\cdot \vec{E}=\langle \alpha^H_0\rangle E^2$,
which allows us to write 
\begin{equation}
\langle \alpha^H_0\rangle E^2 = \int d\vec{X} d\vec{Q}\rho(\vec{X},\vec{Q}) \sum_{H_k} (\vec{E}\cdot \vec{n}_{k})^2 \alpha^{H_k}_0(\vec{X},\vec{Q}),
\label{eq:alpha_H_SM}
\end{equation}
where $\alpha^{H_k}_0=Z_k(\vec{X},\vec{Q})^2/(m_H\Omega_H^2)$. To derive Eq.~(1) of the main text, we average over the direction of the external field. [It is enough to consider only three electric fields along the $x$,$y$, and $z$ directions of the laboratory frame, and sum the three corresponding expressions from Eq.~(\ref{eq:alpha_H_SM}).]

\begin{figure*}[h!]
\begin{center}
\includegraphics[width=\textwidth]{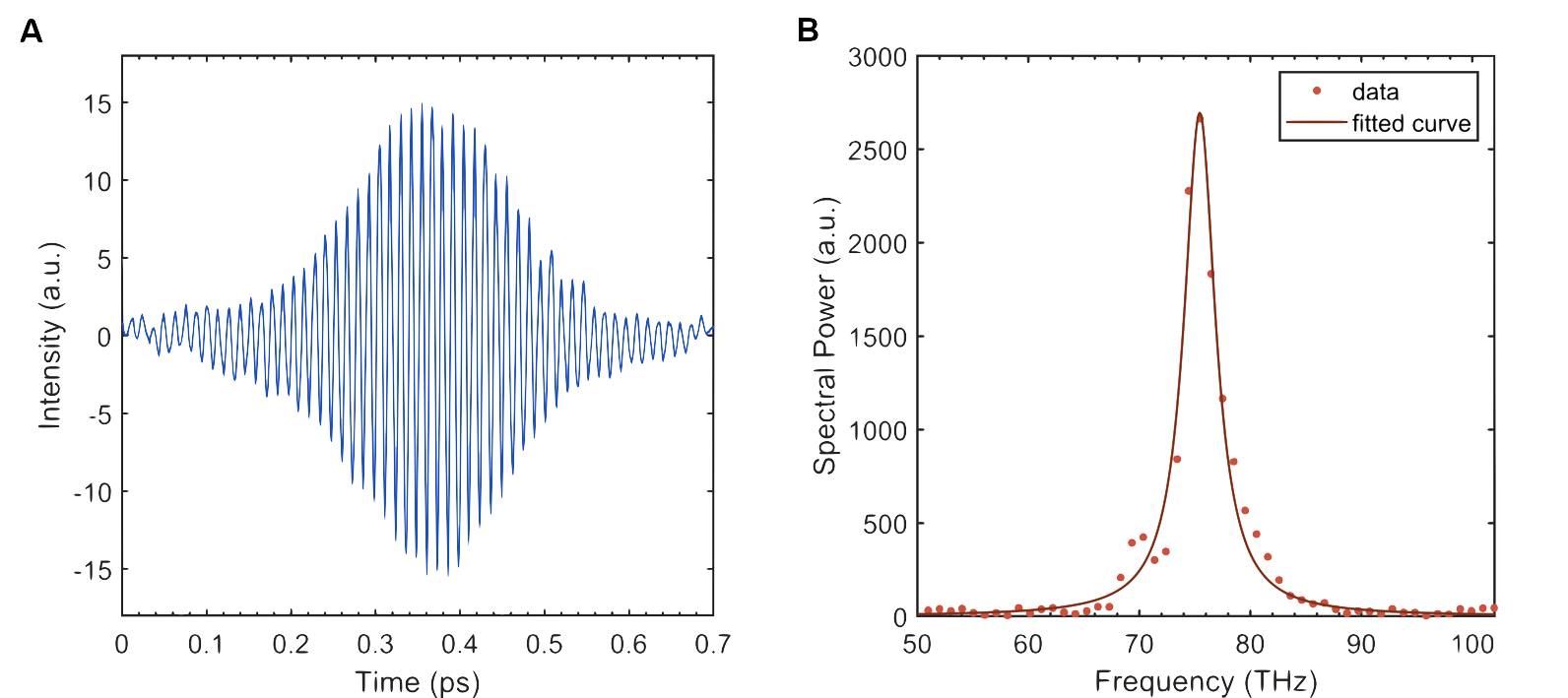}
\end{center}
\caption{(A) Signal as collected from a Michelson interferometer, where the delay between the recombined mid-IR pump pulses is created via use of a translation stage. (B) Result of a fast Fourier transform of the signal in (A), showing clearly the frequency spectrum. The solid line is the Lorentizan fit of the data, from which we extract the central wavelength and full width half maximum (FWHM).}
\label{fig:FTIR}
\end{figure*}

\subsection{Spectroscopy}
In the optical Kerr-effect setup, the probe pulses are near-infrared with wavelength $\lambda=1028$nm and pulse duration $\tau=270$fs; the pump pulses are variable-wavelength mid-IR, inducing birefringence at the probe sample after being delayed inside the test sample. The pump beam is modulated with an optical chopper at 750 Hz, and signal is detected via lock-in detection. The pump pulses are produced by an optical parametric amplifier (Light Conversion Orpheus) coupled to an amplified pulsed laser (Light Conversion Pharos). The wavelength for each measured datapoint is independently calibrated using Fourier-transform infrared spectroscopy (see the inset in Fig.~2B of the Main Text). For this step, the pump pulses and signal are locked-in at 131 Hz. We utilize a simple Michelson interferometer setup to determine the wavelength of the pump probe for each measured datapoint in the experiment. One example of such transform to obtain the wavelength is shown in Fig.~\ref{fig:FTIR}.

The data in Fig.~2B of the main text can also be used to calculate the extinction coefficient, $\kappa$: neglecting the Fresnel losses at the perovskite surface in comparison to bulk absorption, $\kappa$ can be found from the ratio of the intensity transmitted through \textbf{T} to the incident intensity. The ratio of transmission with ($I_{\text{with}}$) and without ($I_{\text{without}}$) the test sample (given that the attenuation of light with distance \(z\) is exponential, $I = \exp{-\frac{4\pi \kappa z}{\lambda_0}}$) is given by
\begin{equation}
    \frac{I_{\text{with}}}{I_{\text{without}}} = \frac{ \exp \left(-4\pi \kappa_{\text{air}} \nicefrac{(z-d)}{\lambda_0} \right) \exp \left( -4\pi \kappa \, \nicefrac{d}{\lambda_0}\right)} { \exp \left(-4\pi \kappa_{\text{air}}\,\nicefrac{z}{\lambda_0} \right) } \qquad \to \qquad  \kappa = -\ln\left[\frac{I_{\text{with}}}{I_{\text{without}}}\right] \times \frac{\lambda_0}{4\pi d},
\end{equation}
where $d$ is the sample thickness, $\lambda_0$ is the wavelength of the pump in vacuum, and we have assumed that the extinction coefficient of air, $\kappa_{\text{air}}$ is 0. 

\subsection{Extinction coefficient}
\begin{figure*}[h!]
\begin{center}
\includegraphics[width=\textwidth]{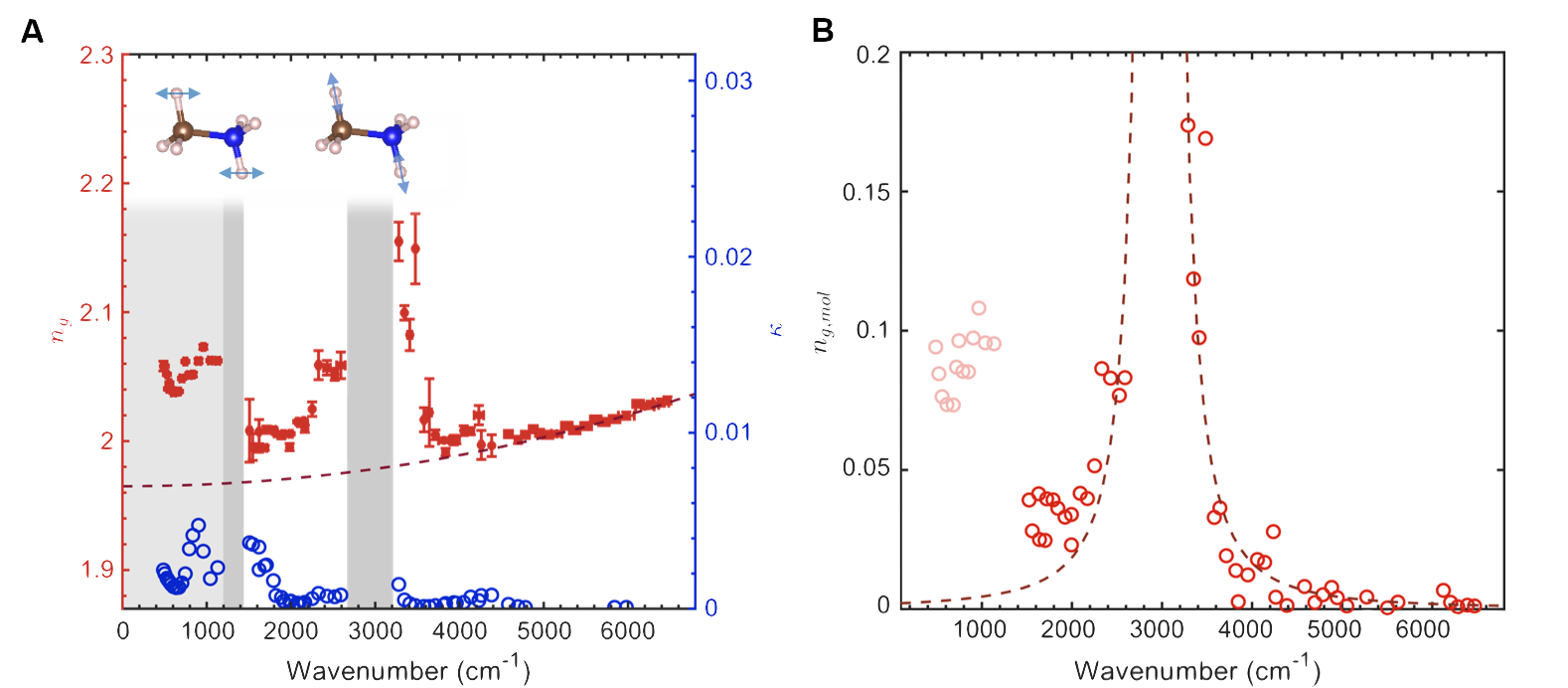}
\end{center}
\caption{Extinction data appended on the data as presented in the main text, see Fig.~3 of the main text. The red dashed line is the electronic group index obtained from the Sellmeier fit to high energy data of Ref.~\cite{ArtemArxiv2020}. The two prominent regions of vibrational resonance are indicated by the grey shading. Cartoon depictions of these families of vibrational modes are presented above the shaded regions, with the higher energy modes indicating C(N)--H stretching and the lower energy modes C(N)-H bending. Only one C(N)-H bond's direction of motion (toward against the bond) is shown with a blue arrow, for clarity, representing the entire cluster of modes in that energy region. (B) Molecular contribution to the group index, \(n_{g,mol}\), around the region of stretching modes in (A) and Fig.~3 of the main text, obtained through subtracting the electronic component from the Sellmeier fit. The dashed line is the resultant fit using Eq. (2) of the main text.}
\label{fig:extinction}
\end{figure*}

Figure~\ref{fig:extinction}A shows the obtained extinction together with the data from Fig.~3 of the main text. The missing values in the regions of resonance are a result of strong absorption of the sample. The low value of $\kappa$ in the region outside the resonance suggests that these modes (clustered near a single resonance energy)  are relatively strong and can be approximately considered in isolation from the other modes. Indeed, upon subtracting the electronic cage contribution to the refractive index from the high-energy Sellmeier fit to our previous data~\cite{ArtemArxiv2020} as depicted in Fig.~\ref{fig:extinction}B, we achieve an excellent fit with Eq.~(2) of the main text, which is derived for a single resonance energy $\hbar\Omega_H$.

\subsection{Derivation of Eq. (2) of the main text}
We use the Lorentz-Lorenz relation between the phase refractive index $n$ and the polarizabilities of all degrees of freedom of the system $\{\alpha_i \}$ \cite{Born2019}:  
\begin{equation*}
    \frac{n_{ph}^2-1}{n_{ph}^2+2} = \sum_i \frac{4\pi N}{3}\alpha^i(\omega),
\end{equation*}
\noindent where $N$ is the number density of atoms and the sum goes over all degrees of freedom including electrons. As can be seen in Fig.~3 of the main text, the molecular contribution to the total phase refractive index $n_{ph}=n_{ph,el}+n_{ph,mol}$ is small. Therefore, $n_{ph,mol}$ and the molecular polarizability $\alpha_{mol}^i$ are related as
\begin{equation}
   \frac{6 n_{ph,el}}{(  n_{ph,el}^2 +2 )^2}  n_{ph,mol} \approx  \sideset{}{'}\sum_{i \in \{ \text{mol}\}}  \frac{4\pi N }{3} \alpha^i_{mol}(\omega),
   \label{eq:taylor}
\end{equation}
\noindent where the sum goes over molecular degrees of freedom. The polarizability of the C(N)--H stretching mode in the vicinity of its resonant frequency $\Omega_H$ diverges as $ \alpha^H_{mol}(\omega)\approx \langle\alpha^H_{0}\rangle {\Omega_H^2}/(\Omega_H^2-\omega^2)$ with $\langle \alpha^H_{0}\rangle$ being the average static (longitudinal) polarizability of the C(N)--H bond. Therefore, we can further simplify Eq.~(\ref{eq:taylor}) by keeping only the term corresponding to the mode in question. 

We convert the phase index $n_{ph,mol}=n_{ph}-n_{ph,el}$ into the group index $n_{g,mol} \approx \, \partial (\omega  n_{ph,mol})/\partial \omega$, and re-write Eq.~(\ref{eq:taylor})
\begin{equation}
    n_{g,mol} (\omega) \approx \langle \alpha^H_{0}\rangle \frac{4\pi N}{3}\frac{(\bar{n}_{el}^2+2)^2}{12\bar{n}_{el}} \cdot \frac{\Omega_H^2}{(\Omega_H-\omega)^2},
\label{eq:group_mol_SI}
\end{equation}
\noindent where $\bar{n}_{el}\equiv  n_{ph,el}(\omega=\Omega_H)$. $N=\frac{1\,\mathrm{molecule}}{(5.92 \text{\r{A}})^3}$ in our calculations (the lattice constant $a=5.92$\r{A} is taken from Ref.~\cite{Weber1978}, see also~\cite{Yin2017,Abia2022}).

\subsection{Uncorrelated dynamics}

If the cage and the molecule are uncorrelated, we write
\begin{equation}
\langle \alpha^H_0 \rangle_{\mathrm{uncorr}} = \frac{1}{3} \sum_{H_k=1}^6 \int  d\vec{X} \, d\vec{Q} \rho_{\mathrm{mol}}(\vec{X})\rho_{\mathrm{cage}}(\vec{Q}) \, \alpha^{H_k}_0(\vec{X},\vec{Q}),
\end{equation}
where $\rho_{\mathrm{mol}}(\vec{X})$ [$\rho_{\mathrm{cage}}(\vec{Q})$] is the density matrix that describes the molecule [the cage]. As there are (by assumption) no correlations between the lattice and MA, we place the molecule in the middle of the lattice unit, and assume that all allowed positions of the C--N and C(N)--H bonds are equally likely. Furthermore, we assume that $\alpha^{H_k}_0(\vec{X},\vec{Q})\simeq \alpha^{H_k}_0(\vec{X},\vec{Q_0})+\vec{c} (\vec{Q}-\vec{Q}_0)$, where $\vec{Q}_0$ are average positions of the atoms in the cage and $\vec{c}$ is the vector of expansion coefficients. [Note that this assumption is not necessary for our derivation as will become clear below.] This assumption is natural, as the atoms in the cage can move only around their equilibrium positions with a relatively small amplitude. Furthermore, it simplifies the derivation because (for harmonic dynamics of the cage) the linear term $(\vec{Q}-\vec{Q}_0)$ disappears.  
 
With this, we write       
\begin{equation}
\langle \alpha^H_0 \rangle_{\mathrm{uncorr}} = \int  d\vec{X} \rho_{\mathrm{mol}}(\vec{X}) \, \alpha^{H_N}_0(\vec{X},\vec{Q}_0)+\int  d\vec{X} \rho_{\mathrm{mol}}(\vec{X}) \, \alpha^{H_C}_0(\vec{X},\vec{Q}_0),
\end{equation}
where $H_N$ ($H_C$) refers to a hydrogen atom from the N--H (C--H) bond.
We assume that the longitudinal polarizability depends only on $\vec{n}\vec{E}_{\vec{Q}_0}(\vec{X}_H)$, i.e.,
 $\alpha^{H}_0(\vec{X},\vec{Q}_0)=\alpha^{H}_0(\vec{n}\vec{E}_{\vec{Q}_0}(\vec{X}_H))$ where $\vec{E}_{\vec{Q}_0}(\vec{X}_H)$ is the crystal field at the locus of the C(N)-H bond. This assumption is based on (i) the observation that the fields in the center of the lattice unit of a cubic lattice are weak (they cancel each other due to the symmetry of the unit), and (ii) on the results of our DFT calculations. Further physical intuition can be gained by using a charged ellipsoid model for the C(N)--H bond.
 
Since the fields in the center of the lattice unit of a cubic lattice are weak, we write
\begin{equation}
\langle \alpha^H_0 \rangle_{\mathrm{uncorr}} = \int  d\vec{X} \rho_{\mathrm{mol}}(\vec{X}) [\alpha^{H_N}_0(\vec E=0)+\alpha^{H_C}_0(\vec E=0)+ C_{H_C} \vec{n}\vec{E}_{\vec{Q}_0}(\vec{X}_{H_C})+C_{H_N} \vec{n}\vec{E}_{\vec{Q}_0}(\vec{X}_{H_N})],
\end{equation}
where the expansion coefficients $C_{H_k}$ do not depend on the electric field. The integral $\int  d\vec{X} \rho_{\mathrm{mol}}(\vec{X}) \vec{n}\vec{E}_{\vec{Q}_0}(\vec{X}_{H})$ vanishes. To see this, we expand the vector $\vec{n}$ in components parallel and perpendicular to $\vec{R}$ (for a sketch of the geometry, see Fig.~1 of the main text). The two resulting integrals $\int  d\vec{X} \rho_{\mathrm{mol}}(\vec{X}) \vec{n}_{par}\vec{E}_{\vec{Q}_0}(\vec{X}_{H})$ and $\int  d\vec{X} \rho_{\mathrm{mol}}(\vec{X}) \vec{n}_{perp}\vec{E}_{\vec{Q}_0}(\vec{X}_{H})$ vanish due to Gauss's law and Kirchhoff's loop rule, respectively, which allows us to derive the expression presented in the main text:
\begin{equation}
\langle \alpha^H_0 \rangle_{\mathrm{uncorr}} = \alpha^{H_N}_0(\vec E=0)+\alpha^{H_C}_0(\vec E=0).
\end{equation}

\subsection{Correlated dynamics}

To estimate the value of the polarizability in case of correlated dynamics, we approximate the density matrix as
\begin{equation}
\rho(\vec{X},\vec{Q})=\rho_{mol}(\vec{X};\vec{Q}_0)\rho_{cage}(\vec{Q}),
\label{eq:SM_correlated_density_matrix}
\end{equation}
where $\rho_{mol}(\vec{X};\vec{Q}_0)$ describes the molecule assuming that the atoms in the cage are located at their average positions, $\vec{Q}_0$; $\rho_{cage}(\vec{Q})$ describes the atoms in the cage. 
Equation~(\ref{eq:SM_correlated_density_matrix}) relies on the observation that the time scale for the cage dynamics is sub-ps whereas the dynamics of the C--N reorientation is `much' longer, on the order of a few ps~\cite{Leguy2015,Quarti2016,Mattoni2017}.

We assume that $\alpha^{H_k}_0(\vec{X},\vec{Q})\simeq \alpha^{H_k}_0(\vec{X},\vec{Q_0})+\vec{c} (\vec{Q}-\vec{Q}_0)$ and that the dynamics of the cage is harmonic, which allows us to write 
\begin{equation}
\langle \alpha^H_0\rangle \simeq \frac{1}{3}\int d\vec{X} \rho_{mol}(\vec{X};\vec{Q}_0) \sum_k \alpha^{H_k}_0(\vec{X},\vec{Q_0}).
\end{equation}
To establish the form of $\rho_{mol}(\vec{X};\vec{Q}_0)$, we note that there are six degenerate positions of the C--N bond in the three (parallel to the surfaces) planes of symmetry of a cube, see, e.g., ~\cite{Leguy2015}. The interaction between the N--H bonds and the halide atoms in the cage determines the positions of the hydrogen atoms~\cite{Lee2016,Motta2016,Yin2017}. For simplicity, we assume that the complex N--H--Br forms a straight line, so that we can directly use our DFT results presented in Fig. 1 of the main text. This assumption is a crude approximation of the molecular position as the C(N)--H bond features a rapid `wobbling-within-a-cone' motion~\cite{Leguy2015}, for review see~\cite{Gallop2018}. However, with this, we can provide an intuitive illustration of the approach (which is the main purpose of our work), and, furthermore, to put a lower bound on the electric field felt by a molecule.  Finally, we derive
\begin{equation}
\langle \alpha^H_0\rangle=\frac{1 }{3 m_H\Omega_H^2} \sum_k Z^2_k(\vec{E}_{\vec{Q}_0}(\vec{X}_{H_k})\cdot \vec{n}_k),
\end{equation}
where $\vec{E}_{\vec{Q}_0}(\vec{X}_{H_i})$ is the crystal field at the locus of the C(N)--H bond. Using this expression and Fig.~1 of the main text, we estimate $\vec{E}_{\vec{Q}_0}(\vec{X}_{H_i})$ whose value is reported in the main text. 
[To include the uncertainty in the position of $\Omega_H$ one could use $e^2/(m_H\Omega_H^2)$ in the range $0.41-0.44$ \r{A}$^3$.]

\subsection{Coulomb fields inside the lattice}
In HOIPs, molecules are located close to the centers of lattice units, and experience weak electric fields. 
To estimate these fields, we calculate the electric potential close to the center of a lattice unit assuming that Br and Pb are point particles with charges -1|e| and +2|e|, respectively.   
 This potential can be calculated from the Taylor series whose first terms read
 \begin{equation}
  \label{eq:SM:Taylor_Coulomb}
 \begin{split}
 \phi(\vec{x})=\phi(0)+e \vec{x}\sum_{Br_k}\frac{\vec{Q}_{Br_k}}{|\vec{Q}_{Br_k}|^3}-2e \vec{x}\sum_{Pb_k}\frac{\vec{Q}_{Pb_k}}{|\vec{Q}_{Pb_k}|^3}-e\sum_{Br_k}\frac{3(\vec{Q}_{Br_k} \vec{x})^2-x^2 Q^2_{Br_k}}{|\vec{Q}_{Br_k}|^5} \\
 +2e\sum_{Pb_k}\frac{3(\vec{Q}_{Pb_k} \vec{x})^2-x^2 Q^2_{Pb_k}}{|\vec{Q}_{Pb_k}|^5} + ... \;,
 \end{split}
 \end{equation}
where $\vec{Q}_{Br_k}$ ($\vec{Q}_{Pb_k}$) is the position of the $k$th Br (Pb) atom, and the sum goes over all atoms of a lattice unit. Due to the cubic symmetry of the lattice, the sums presented in Eq.~(\ref{eq:SM:Taylor_Coulomb}) vanish,  and one needs to consider higher-order terms in the expansion, which lead to weak electric fields for small values of $x$.

From the symmetry argument, one can expect that the electric fields behave as $|\vec{E}|\sim e x^3/(a/2)^5$ close to the center of the unit cell, here $a=5.92$\r{A} is the lattice constant~\cite{Weber1978}. [It is worthwhile noting that a fast decay of the electric field, $\sim 1/|\vec{Q}|^5$ is the reason why it is enough to consider a single lattice unit for our estimation.] Assuming that $x\simeq 1$\r{A}, we obtain $|\vec{E}|\sim 0.06$ V/\r{A}. [Numerical calculations give even a smaller value, and confirm the order of magnitude of this estimate.] This value is too small to account for the measured value of the polarizability. Therefore, we conclude that one cannot 
consider ions in the cage as point-like objects. The simplest extension consistent with cubic symmetry of the lattice is a quadruple instead of Pb--Br--Pb complex. This leads to the potential $\phi\sim \frac{D}{ 2 d^3}$ at the locus of the N--H bond, here $D$ is the quadrupolar moment and $d=$2.5\r{A} (see the main text). We put an estimate $D\in [0.5,1.6]$~e\r{A}$^2$ -- a natural atomic value -- by comparing to the electric field deduced from renormalization of the polarizability. In the main text, we report only the central value.

\subsection{Hydrogen Bond}
 We estimate H-bonding interactions from the energy required to move the molecule to the center of the cage.  Using our estimate of $D$, we find that this energy per a hydrogen bond is $Z D/(2d^3)-Z D/(2d_1^3)\sim 0.1$~eV, where  $d_1\simeq 3.2$\r{A}, and we use the Born charge $Z=0.4$e to represent the charge of the N--H hydrogen, see Fig.~1 of the main text. Although, this estimate includes a number of approximations, it agrees well with the values reported in the literature, supporting our conclusion regarding usefulness of bond spectroscopy.

\bibliography{bibli}

\end{document}